# Three-dimensional Burgers' Equation as a Model for the Large-Scale Structure Formation in the Universe *

Sergei F. Shandarin

July 5, 1995

**Abstract**

As galaxy redshift surveys probe deeper into the universe, they uncover ever more dramatic structures in the large-scale distribution of galaxies. In particular, the CfA2 and SSRS2 surveys to an apparent magnitude limit of 15.5 exhibit an impressive complex of sheets, filaments, and clusters. The formation of the large-scale structure in the universe results from the gravitational amplification of the primordial small perturbations of density. The primordial density perturbations are thought to be random fields originated as quantum fluctuations at the very early stage. Thus the understanding of the formation of the large-scale structure may reveal important information about the early universe and the laws of fundamental physics. One of the major obstacles to understanding the formation of the large-scale structure is the complexity of the evolution of the density inhomogeneities at the nonlinear stage when the observable structures form. One way of addressing this problem is to run three-dimensional numerical simulations. Here we review another approach based on the approximate analytic model of the nonlinear gravitational instability utilizing Burgers' equation of the nonlinear diffusion.

## 1 Introduction

The term large-scale structure in the universe is referred to a distribution of galaxies on the scales roughly from 1 $Mpc$ to 100 $Mpc$ [17], where 1 $Mpc = 10^6$ $pc \approx 3 \cdot 10^{24}$ $cm$ is a unit of length commonly used in cosmology. Galaxies can not probe much smaller scales because of discreteness, and on larger scales the galaxy distribution becomes almost homogeneous. The redshift surveys reveal spectacular abundance of structures often described as filamentary, network, or bubble structure [17], [6], [4] (see Fig.1). The origin of the large-scale structure is one of the most important problems in modern cosmology. Many fundamental issues in physics, cosmology and astronomy ranging from speculations on the physical nature of dark matter, to the measurement of angular anisotropy of the microwave background radiation and determination of the epoch of galaxy formation join together here [21].

---

The most popular and best developed class of theories of structure formation is based on the assumption that it started from primeval small amplitude density perturbations which grew by gravitational instability. Primeval perturbations are assumed to arise as vacuum fluctuations during the very early stage when the universe was expanding exponentially (inflationary universe) [13]. Afterwards, the density perturbations had a long history before they become galaxies, clusters of galaxies, superclusters and voids. The formation of galaxies is a very difficult problem itself. Many complex physical processes like star formation and supernova explosions are very important for understanding the galaxy formation. We shall discuss the mass distribution assuming that galaxies are fairly good (though not perfect) tracers of mass on large scales.

As long as the density perturbations are small by amplitude their evolution is described by the linear theory of gravitational instability (see e.g. [28]; [18]; [22]. The linear theory is very well understood; in particular, it predicts the rate of growth of the perturbations at different stages of the evolution of the universe. Unfortunately, it is not applicable to the nonlinear stage when the amplitude of the density perturbations becomes large and the observable structures (sheets, filaments, and clusters of galaxies) form.

One of the most difficult obstacles arising in understanding models is the analysis of the evolution of perturbations at the nonlinear stage, when the typical amplitudes of density inhomogeneities become larger than mean density in the universe: $\delta\rho/\rho \geq 1$. The most straightforward way to address the problem of the nonlinear evolution is to do gravitational three-dimensional N-body simulations. Usually in simulations of this type the medium is assumed to consist of collisionless particles, in agreement with the popular hypothesis that the most of the mass in the universe is in the form of weakly interacting particles such as massive neutrinos or axions. The trajectory of each particle is calculated in the gravitational field generated by all the particles. Boundary conditions are commonly assumed to be periodic.

Here we review an alternative approach to the problem of the large-scale structure in the universe. We present an approximate analytic technique to solve the nonlinear gravitational instability based on Burgers' equation. Other analytic and semi-analytic methods are discussed in an excellent review by Sahni and Coles [19].

The initial condition is the result of the linear theory applied to the earlier stages. In the linear regime the density fluctuations are assumed to be a Gaussian random field specified by the spectrum and the amplitude. The current measurements of the angular fluctuations in the temperature of the microwave background radiation by COBE (Cosmic Background Explorer) and other experiments put strong constraints on the initial fluctuations. In particular, the amplitude of the temperature fluctuations suggests that the scale where the density fluctuations have recently reached nonlinearity is about $5 - 10$ $Mpc$ which is in a good agreement with the observations of the large-scale galaxy distribution (see e.g. [26]).

According to the prediction of cosmic inflation, the geometry of the universe is assumed to be flat (see e.g.[13]. For simplicity we also assume that the cosmological term $\Lambda = 0$. Such a model is specified by only two parameters: the Hubble constant $H_0$ describing the current rate of expansion of the universe and the fractional density of baryons $\Omega_b$. The astronomical measurement of the Hubble constant is not very accurate: $50 < H_0 < 100$ $km$ $s^{-1}$ $Mpc^{-1}$; this uncertainty is usually expressed in terms of a parameter $h$:

$H_0 = 100 \ h \ km \ s^{-1} \ Mpc^{-1}$ with $0.5 < h < 1$. The density of baryons as well as the density of other components is convenient to measure in the units of the critical density: $\Omega_b = \bar{\rho}_b/\rho_{cr}$, where the critical density $\rho_{cr} \equiv 3H_0^2/8\pi G \approx 2 \cdot 10^{-29} g \ cm^{-3}$ is a parameter separating the closed cosmological models $\Omega_{tot} > 1$ from the open ones $\Omega_{tot} < 1$. An open universe has negative spatial curvature and expands forever, and a closed universe has positive curvature and eventually collapses. If $\Omega_{tot} = 1$ the spatial geometry of the universe is flat and it expands forever.

We shall assume that about 90% or more of mass in the universe is in a form of weakly interacting collisionless particles (dark matter) and the remaining few percent of the mass is in baryons; $\Omega_{tot} = \Omega_{dm} + \Omega_b = 1$ [21]. Although all luminous objects (e.g. stars) are made from baryons the baryon component dynamically is not very important on large scales. Therefore we shall study the evolution of density fluctuations in the collisionless medium neglecting the baryonic component.

## 2 Basic Equations

The evolution of density inhomogeneities can be described by a system of three partial differential equations comprising the continuity, Euler, and the Poisson equation (see e.g. [28], [18], [22]). It is convenient to exclude the homogeneous expansion of the universe from the analysis by introducing new variables. Instead of the coordinates $\mathbf{r}$ the comoving coordinates $\mathbf{x} = \mathbf{r}/a(t)$ are commonly used, here $a(t)$ is the scale factor describing the homogeneous expansion of the universe; in a flat universe $a(t) \propto t^{2/3}$. Instead of the velocity $\mathbf{u}$ the so called peculiar velocity $\mathbf{u}_p = \mathbf{u} - \dot{a}/a \cdot \mathbf{r}$ is used; if $\mathbf{u}_p = 0$ the velocity field on the scales smaller than the cosmological horizon ($R_H \approx 3{,}000 \ h^{-1} \ Mpc$) is described by the Hubble law: $\mathbf{u} = \frac{\dot{a}}{a} \cdot \mathbf{r} \equiv H(t) \cdot \mathbf{r}$). The velocities and gravitational field in the process we discuss are nonrelativistic ($v \ll c$, $\varphi \ll c^2$) thus the evolution of density inhomogeneities can be described by the equations of classic hydrodynamics and gravity. In terms of the comoving coordinates and peculiar velocities the equations are as follows: the continuity equation

$$\frac{\partial \rho}{\partial t} + \frac{1}{a} \nabla \cdot (\rho \mathbf{u}_p) = -3\frac{\dot{a}}{a}\rho, \qquad (1)$$

the Euler equation

$$\frac{\partial \mathbf{u}_p}{\partial t} + \frac{1}{a}(\mathbf{u}_p \cdot \nabla)\mathbf{u}_p = -\frac{1}{a}\nabla\phi - \frac{\dot{a}}{a}\mathbf{u}_p, \qquad (2)$$

and the Poisson equation

$$\frac{1}{a^2}\nabla^2\phi = 4\pi G(\rho - \overline{\rho}), \qquad (3)$$

where $\dot{a} \equiv da/dt$; $\rho$ and $\overline{\rho}$ are respectively the density and mean density of mass; $\phi$ is the perturbation of the gravitational potential due to the inhomogeneities of density; $G$ is the gravitational constant.

The pressure is neglected since we study the medium interacting only gravitationally. Additional terms on the right hand side of the continuity and Euler equations are due to the homogeneous expansion of the universe and the factor $1/a$ is due to differentiation with respect to the comoving coordinates $\mathbf{x}$: $\nabla \equiv \partial/\partial x_i \equiv a \cdot \partial/\partial r_i$. We assume that the initial condition is small density and smooth velocity perturbations imposed on a homogeneous density distribution.

As long as the amplitude of the density perturbations remains small their evolution can be analyzed in the linear approximation obtained by the linearization of the above equations. The exact solution of the linearized system has one growing mode which is the major object of our analysis. The velocity in the growing mode is a potential vector field. In the linear regime the spatial structure of the perturbations (in the comoving coordinates) remains unchanged and its amplitude is proportional to the growing solution $b_g$ of the differential equation

$$a\frac{d^2b}{dt^2} + 2\dot{a}\frac{db}{dt} + 3\ddot{a}b = 0. \tag{4}$$

The scale factor $a$ is assumed to be a known function of time and in a flat matter dominated universe $b_g(t) \propto a(t) \propto t^{2/3}$. It is convenient to make yet another transformation of the variables:

$$\eta(\mathbf{x}, b_g) = a^3 \rho(\mathbf{x}, t), \tag{5}$$

$$\mathbf{v}(\mathbf{x}, b_g) \equiv \frac{d\mathbf{x}}{db_g} = \frac{1}{a\dot{b}_g} \mathbf{v}_p(\mathbf{x}, t), \tag{6}$$

$$\varphi(\mathbf{x}, b_g) = (3\ddot{a}ab_g)^{-1} \phi(\mathbf{x}, t), \tag{7}$$

and also to reparametrize the time coordinate by the monotonic function of time $b_g(t)$ describing the growth of the perturbations. Finally, after explicit use of the function $b_g = a$ (which assumes $\Omega_{tot} = 1$) and introducing the velocity potential $\Phi$: $\mathbf{v} = -\nabla\Phi$ the equations take the form

$$\frac{\partial \eta}{\partial a} + \nabla \cdot (\eta \mathbf{v}) = 0, \tag{8}$$

$$\frac{\partial \mathbf{v}}{\partial a} + (\mathbf{v} \cdot \nabla)\mathbf{v} = -\frac{3}{2a}\nabla(\varphi - \Phi), \tag{9}$$

$$\nabla^2 \varphi = \frac{1}{a}\frac{\eta - \bar{\eta}}{\bar{\eta}}. \tag{10}$$

In the linear regime both the gravitational potential $\varphi$ and velocity potential $\Phi$ remain constant and equal to each other, and the right hand side of eq.9 vanishes. In 1970 Zel'dovich [27] suggested to extrapolate this condition well into the nonlinear regime; the corresponding solution is known in cosmology as the Zel'dovich approximation.

## 3 Zel'dovich Approximation

The Zel'dovich approximation is convenient to formulate as a mapping from the Lagrangian space $L\{q\}$ into Eulerian space $E\{x\}$

$$\mathbf{x}(\mathbf{q}, a) = \mathbf{q} + a \cdot \mathbf{v}_0(\mathbf{q}), \tag{11}$$

which obviously follows from eq.9 assuming that $\varphi \approx \Phi$. The initial velocity potential $\Phi_0(\mathbf{q})$ ($v_{0i}(\mathbf{q}) = -\partial\Phi_0/\partial q_i$) is assumed to be a smooth random Gaussian field specified by the spectrum $P_\Phi(k)$. In cosmology the initial condition is usually characterized by the spectrum $P_\delta(k)$ of the linear density fluctuations $\delta \equiv (\rho - \bar\rho)/\bar\rho \equiv (\eta - \bar\eta)/\bar\eta$ which is obviously related to the spectrum of the potential

$$P_\Phi(k) = k^{-4}\, P_\delta(k). \tag{12}$$

Utilizing the law of mass conservation one finds the density as a function of the time coordinate $a$ and the Lagrangian coordinate $\mathbf{q}$

$$\eta(\mathbf{q}, a) = \frac{\bar\eta}{[1 - a\,\lambda_1(\mathbf{q})][1 - a\,\lambda_2(\mathbf{q})][1 - a\,\lambda_3(\mathbf{q})]}, \tag{13}$$

where $\lambda_1(\mathbf{q})$, $\lambda_2(\mathbf{q})$ and $\lambda_3(\mathbf{q})$ are the eigenvalues of the deformation tensor $d_{ij} = \partial^2 \Phi_0 / \partial q_i \partial q_j$. Combining eqs.11 and 13 one can find the density distribution in the Eulerian space. It follows from eq.11 and eq.13 that the first objects form at the maxima of the largest eigenvalue and have very oblate shapes. After Zel'dovich they are known in cosmology as "pancakes". Recent three-dimensional gravitational N-body simulations are in a perfect agreement with this conclusion [23]. The pancakes originate as the three-stream flow regions bounded by caustics, the surfaces of formally infinite density. The shape and other characteristics of the pancakes are determined by catastrophe theory [1]. At the later stages the Zel'dovich solution predicts several different types of singularities which are classified in [1].

We use this opportunity to remark that the Zel'dovich approximation (in two-dimensional space) is very similar to the equations describing the propagation of light in geometric optics [29].

The Zel'dovich approximation proved to be very good until orbit crossing when caustics form and the multi-stream flows occur (see e.g. [22] and references therein). (At this stage the original equations must be obviously modified in order to incorporate the multi-stream flows.) However, the Zel'dovich approximation predicts the multi-stream flow regions to broaden very fast which contradicts to the results of the N-body simulations [5], [9]. Numerical studies of the orbits in the multi-stream flow regions show that the velocity component orthogonal to the pancakes randomizes very quickly [12]. As a result the pancakes observed in the N-body simulations remain quite thin. This result has become a physical basis for the adhesion approximation.

## 4  Adhesion Model

The general idea of the adhesion model is very simple. We wish to use the Zel'dovich solution everywhere except the regions of multi-stream flows. By adding a diffusion term into the Euler equation one can suppress the formation of the multi-stream flow regions. Assuming that the gravitational potential is approximately equal to the velocity potential $\varphi \approx \Phi$ and adding a viscosity term $\nu \nabla^2 \mathbf{v}$ in eq.9 one obtains the equation of the nonlinear diffusion [7], [8]

$$\frac{\partial \mathbf{v}}{\partial a} + (\mathbf{v} \cdot \nabla)\mathbf{v} = \nu \nabla^2 \mathbf{v}. \tag{14}$$

Generally speaking the viscosity term need not to be in the form of eq.14 but choosing this particular form one obtains Burgers' equation that has an exact analytic solution [3]. For potential motion $\mathbf{v} = -\nabla \Phi$ eq.14 can be solved by performing the Hopf-Cole substitution $\Phi(\mathbf{x}, a) = -2\nu \log U(\mathbf{x}, a)$. As a result eq.14 translates into the familiar linear diffusion equation

$$\frac{\partial U}{\partial a} = \nu \nabla^2 U. \tag{15}$$

Solving eq.15 for the velocity we obtain

$$\mathbf{v}(\mathbf{x}, a) = \frac{\int d^3q \left(\frac{\mathbf{x}-\mathbf{q}}{a}\right) \exp[-S(\mathbf{x}, a; \mathbf{q})/2\nu]}{\int d^3q \exp[-S(\mathbf{x}, a; \mathbf{q})/2\nu]}, \tag{16}$$

where the "action"

$$S(\mathbf{x}, a; \mathbf{q}) = -\Phi_0(\mathbf{q}) + \frac{(\mathbf{x} - \mathbf{q})^2}{2a}. \tag{17}$$

In cosmology the adhesion model has been used in two forms: one assumes a small but finite value of the viscosity parameter $\nu$ and the other assumes it is infinitesimal: $\nu \to 0$ [19].

For finite $\nu$ the trajectory of a particle can be determined by solving the integral equation [24], [25], [16], [15]

$$\mathbf{x}(\mathbf{q}, a) = \mathbf{q} + \int_0^a da' \, \mathbf{v}[\mathbf{x}(\mathbf{q}, a'), a'], \tag{18}$$

and the resulting density can be determined from the continuity equation

$$\eta(\mathbf{x}, a) = \bar{\eta}/det(\frac{\partial x_i}{\partial q_j}). \tag{19}$$

For an infinitesimal value of the viscosity parameter $\nu \to 0$, the integrals in eq.16 can be evaluated using the method of steepest descents [7], [8], [10], [11]. In this case

$$\mathbf{v}(\mathbf{x}, a) = \frac{\mathbf{x} - \mathbf{q}(\mathbf{x}, a)}{a}, \tag{20}$$

where $\mathbf{q}(\mathbf{x}, a)$ is the coordinate of the absolute minimum of the action $S(\mathbf{x}, a; \mathbf{q})$ at given $\mathbf{x}$ and $a$. The points $\mathbf{q}$ that minimize the action obviously satisfy the Zel'dovich equation 11.

This solution (eq.20) has an interesting geometrical interpretation (see Fig.2). One can find the Eulerian coordinate $\mathbf{x}$ of the particle with initial Lagrangian coordinate $\mathbf{q}$ at a chosen time simply by projecting the apex of the paraboloid

$$P(\mathbf{x}, a; \mathbf{q}) = \frac{(\mathbf{x} - \mathbf{q})^2}{a} + C, \tag{21}$$

assuming that there is a constant $C$ satisfying simultaneously two conditions: 1) the paraboloid $P$ is tangent to the initial velocity potential $\Phi_0$ at $\mathbf{q}$ and 2) it does not cross $\Phi_0$ at any point. At early times $a$ is small and the curvature of the paraboloid is greater than that of $\Phi_0$ and the above two conditions can be easily fulfilled. The Zel'dovich approximation is universally valid at this stage.

As time passes the curvature of the paraboloid decreases. As a result the points **q** appear that do not satisfy the above two conditions. Such a point has been stuck into a surface due to orbit crossing. Again, such surfaces can be found by projecting the apex of the paraboloid but this time the paraboloid is tangent to the initial velocity potential in two points simultaneously. These surfaces correspond to the pancakes; their thickness is proportional to the value of the viscosity parameter and therefore is infinitesimal as $\nu \to 0$.

Later two more types of structures form: the filaments and knots. The filaments correspond to the case when the paraboloid touches the initial potential in three points simultaneously and the knots when it touches the potential in four points. The set of surfaces, filaments and knots make a cellular structure: large regions of low density are separated by the surfaces, the filaments are at the intersections of the surfaces, and the knots are at the intersections of the filaments. This geometrical construction can be viewed as the skeleton of the real structure. In one-dimensional case illustrated by Fig.2 obviously there are no surfaces nor filaments.

## 5 Accuracy of the Adhesion Model

Apart from the regions of high density where the viscosity term (eq.14) plays a significant role the adhesion model is exact in one-dimensional case. It means that the velocity field outside the high density regions is predicted exactly. Also the motion of the clumps is described very accurately. This is because the Zel'dovich solution is exact outside the multi-stream flow regions in one-dimensional case [22].

In more interesting two- and especially three-dimensional case the adhesion model is only an approximation. Therefore the question arises about the accuracy of the model. The both variants of the adhesion model have been thoroughly tested against the gravitational N-body simulations in two and three dimensions. Both the N-body simulation and the adhesion model used the identical initial conditions and were compared at several stages of the evolution.

The geometrical version of the adhesion model was tested against the two-dimensional N-body simulations with the initial power law spectra $P_\delta(k) \propto k^n$ with spectral indices $n = 2, 0$ and $-2$ and various cutoffs [11]

The N-body simulations used the particle-mesh code with $512^2$ particles on equal mesh and periodic boundary conditions (for the details see [2]). The code constructing the skeleton of the structure is described in [10]. It has been found that the skeleton reproduces the density distribution extremely well (see Fig.3) for all choices for the parameters of the initial spectra until the stage when the scale of the nonlinearity $k_{nl}^{-1}$ defined by the equation

$$a^2 \int_0^{k_{nl}} P_\delta(k) \, d^D k = 1 \qquad (22)$$

reaches the characteristic scale of the initial velocity potential $\Phi_0$: $k_{nl}^{-1} \leq R_{\Phi_0}$, here $D$ is the dimensions of the space. The scale of the potential is defined from the expansion of the correlation function $\xi_{\Phi_0}(r) = \xi_{\Phi_0}(0)(1 - r^2/2R_{\Phi_0}^2 + \cdots)$:

$$R_{\Phi_0} = (2D)^{1/2} \frac{\sigma_0}{\sigma_1}, \qquad (23)$$

where $\sigma_0$ and $\sigma_1$ are the dispersions of the potential and its gradient, respectively. In the N-body simulations we deal with finite ranges and therefore $R_{\Phi_0}$ always exists. At the later stages $k_{nl}^{-1} \geq R_{\Phi_0}$ the adhesion model remain qualitatively correct though its accuracy somewhat deteriorates.

The version of the adhesion model utilizing a finite viscosity parameter $\nu$ has been quantitatively compared to fully nonlinear, numerical three-dimensional gravitational N-body simulations [15]. The initial perturbations were again random Gaussian fields with power-law spectra with the spectral indices $n = -2, -1, 0, +1$. The particle-mesh N-body and adhesion simulations both used $128^3$ particles on a $128^3$ mesh and periodic boundary conditions. In these simulations of the adhesion model, the smallest value of the viscosity parameter $\nu$ that did not produce numerical overflows has been used. For further discussion of the N-body and adhesion simulations see [14] and [25] respectively. The both codes calculated the particle positions therefore the corresponding density distributions could be easily generated (Fig.4). The primary tool of the quantitative comparison was the cross-correlation coefficient for the density fields obtained from the gravitational N-body simulation and the adhesion model with the identical initial conditions. Also the density distribution functions and the power spectra were compared. The adhesion model produces an excessively filamentary distribution due to smoothing effects in high density regions. As a result the density distribution function in the adhesion model is lower than that of the N-body simulation at high densities $\rho/\bar{\rho} \geq 10 - 20$ depending on the initial spectrum and the power spectrum of the nonlinear distribution falls off steeper on small scales $k \geq k_{nl}$.

# 6 Summary

The adhesion model based on three-dimensional Burgers' equation of the nonlinear diffusion has been found to work very well in explaining the large-scale features of the structure of the universe. Comparison with gravitational two- and three-dimensional N-body simulations has shown remarkable agreement till very late nonlinear times. The main drawback of the adhesion approximation is probably the fact that it cannot describe accurately the density distribution within pancakes and filaments.

The adhesion model provides a natural qualitative explanation of the origin of the large-scale coherent structures such as superpancakes and superfilaments, as a result of coherent motion of clumps due to large-scale inhomogeneities in the initial gravitational potential (compare Fig.1 and 5). The formation and evolution of large-scale structure is described by the adhesion model as a two stage process [20]. During the first stage matter falls into pancakes and then moves along them towards filaments and then along filaments to collect finally in knots. At the end of the first stage the formation of the skeleton of the large scale structure is complete and virtually all of the matter in the universe is located in one of three structural units: pancakes, filaments or knots. The second stage sees the deformation of the large-scale structure skeleton due to the dynamical motion of pancakes, filaments and especially knots. At this stage knots merge into larger knots and small voids disappear giving space to growth of the larger ones. Eventually almost all the mass concentrates in knots. Depending on the initial spectrum the knots may move coherently in such a manner that they concentrate to superpancakes and superfilaments. The superpancakes and superfilaments can be identified by applying the adhesion model to *smoothed* initial potential [11].

**Acknowledgment:** I am grateful to my collaborators on the adhesion model S.Gurbatov, L.Kofman, D.Pogosyan, V.Sahni, A.Saichev, and B.Sathyaprakash for innumerous discussions. I acknowledge NSF grant AST-9021414, NSF EPSCoR grant OSR-9255223, and NASA grant NAGW-2923.

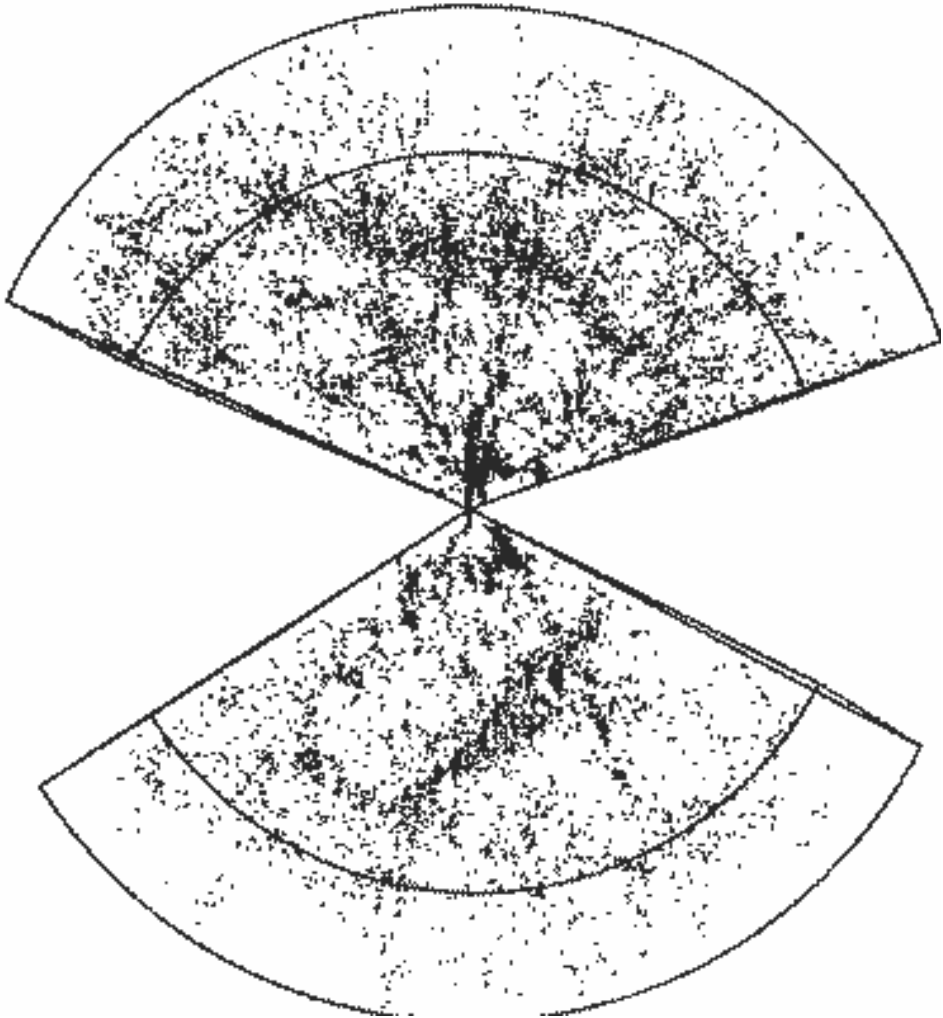

Figure 1: This is a low resolution version of Fig.2 from [4]. It displays galaxies from two redshift catalogs to an apparent magnitude limit of 15.5 and distances less than 120 $h^{-1}$ $Mpc$: CfA2 (north; upper portion) and SSRS2 (south; lower portion). For CfA2 the box shows the declination limits $8°.5 \leq \delta \leq 44°.5$ and the right ascension limits $8^h \leq \alpha \leq 17^h$. In the south the box limits are $-40° \leq \delta \leq -2°.5$ and the right ascension limits are $20^h.8 \leq \alpha \leq 4^h$. There are 9325 galaxies in the original image.

Figure 2: Geometrical prescription of descending a paraboloid onto the initial velocity potential in order to find the Eulerian positions of particles and knots in one-dimensional case. The particle having Lagrangian coordinate $q_0$ in the uppermost panel has the Eulerian coordinate of the paraboloid apex. In the middle panel corresponding to a later stage it is stuck into the knot $m_1$ which velocity is shown by the arrow. The knot $m_2$ is determined by the second paraboloid in the middle panel. The lower panel shows the knot $M$ formed as a result of merging $m_1$ and $m_2$. Adapted from [20].

Figure 3: Composite picture of the results of the two-dimensional N-body simulation and the adhesion model for the flat initial spectrum $P_\delta(k) \propto k^0$. The left hand side panel shows the model with the cutoff of the initial spectrum at $k_c = 32\ k_f$, here $k_f$ is the fundamental frequency corresponding to the box size: $L_{box} = 2\pi/k_f$; the right hand side panel shows the model without a cutoff (except at the Nyquist frequency $k_{Ny} = 256$). Not all the particles can be shown, but this is a fair sample). Solid lines and circles represent the skeleton of the structure constructed by the paraboloid technique. The area of circles is proportional of the mass of knots. Adapted from [11].

Figure 4: A gray scale plot of thin ($L_{box}/128$) slices through the simulation cubes for $n = 0$ (left hand side panels) and $n = -1$ (right hand side panels) initial spectra, at the stages when $k_{nl} = 8k_f$. The gravitational three-dimensional N-body simulations are shown in the bottom panels and the adhesion model simulations are shown in the top panels. Adapted from [15].

Figure 5: Simulated galaxy distributions are drawn from the adhesion model simulation based on the biased Cold Dark Matter cosmological model. These redshift-angle projections show all "galaxies" with apparent magnitude less than 16.5 and distance less than 180 $h^{-1}Mpc$. Adapted from [24].